\documentclass[preprint,12pt]{elsarticle}
\usepackage{xcolor}

\usepackage{amssymb}
\usepackage{algorithm}
\usepackage{algpseudocode}
\usepackage{hyperref}

\journal{Journal of Modelling in Management (accepted)}

\begin{document}

\newcommand{\red}[1]{{\textcolor{red}{#1}}}
\newcommand{\blue}[1]{{\textcolor{blue}{#1}}}
\newcommand{\orange}[1]{{\textcolor{orange}{#1}}}
\newcommand{\green}[1]{{\textcolor{green}{#1}}}
\newcommand{\R}{\mathbb{R}}
\newcommand{\N}{\mathbb{N}}
\newcommand{\RP}{\mathbb{R}^+}
\newcommand{\NP}{\mathbb{N}^+}

\begin{frontmatter}

\title{Optimising task allocation to balance business goals and worker well-being for financial service workforces}

\author[inst1]{Christopher Duckworth}

\affiliation[inst1]{organization = {School of Electronics and Computer Science, University of Southampton},%Department and Organization
            %addressline = {Address One}, 
            city = {Southampton},
            %postcode = {}, 
            %state = {Southampton},
            country={UK}}

\author[inst1]{Zlatko Zlatev}
\author[inst2]{James Sciberras}
\author[inst2]{Peter Hallett}
\author[inst1]{Enrico Gerding}

\affiliation[inst2]{organization={OnCorps},%Department and Organization
            %addressline={Address Two}, 
            city={Bristol},
            %postcode={}, 
            %state={Bristol},
            country={UK}}

\begin{abstract}

\subsection*{Purpose}
Financial service companies manage huge volumes of data which requires timely error identification and resolution. The associated tasks to resolve these errors frequently put financial analyst workforces under significant pressure leading to resourcing challenges and increased business risk. To address this challenge, we introduce a formal task allocation model which considers both business orientated goals and analyst well-being. 

\subsection*{Methodology}
We use a Genetic Algorithm (GA) to optimise our formal model to allocate and schedule tasks to analysts. The proposed solution is able to allocate tasks to analysts with appropriate skills and experience, while taking into account staff well-being objectives. 

\subsection*{Findings}
We demonstrate our GA model outperforms baseline heuristics, current working practice, and is applicable to a range of single and multi-objective real-world scenarios. We discuss the potential for metaheuristics (such as GAs) to efficiently find sufficiently good allocations which can provide recommendations for financial service managers in-the-loop.

\subsection*{Originality}
A key gap in existing allocation and scheduling models, is fully considering worker well-being. This paper presents an allocation model which explicitly optimises for well-being while still improving on current working practice for efficiency.

\end{abstract}


% \begin{highlights}
% \end{highlights}

\begin{keyword}
workforce \sep task allocation \sep genetic algorithm \sep financial services
\end{keyword}

\end{frontmatter}


\section{Introduction}
\label{sec:intro}

Financial service companies acquire huge volumes of data which requires timely error checking and resolution. Specifically, asset managers must accurately publish the per-share value (Net Asset Value; NAV) of mutual or exchange-traded funds at the end of each trading day. Publication of the NAV is reliant on accurate trading data with errors leading to misreporting, significant fines, and, ultimately, large-scale commercial impacts for asset managers \citep{oncorps2024,funds-europe2024}. Error identification and resolution is often a time consuming and overly manual set of tasks. As data volume increases, companies are finding it increasingly hard to monitor all data streams effectively, placing their workforces under increasing pressure.

The human-led checking, analysis, and correction of large volumes of data has not only increased business risk but has had a detrimental impact on staff well-being. Financial Analysts report unsustainably long workhours \cite{labor2023financial} and detrimental impacts on physical and mental health \cite{goldman2022internal, wso2021}, which have led to voluntary staff turnover rates rising over the last decade to be over 20\% annually \cite{labor2022us}. Automation and optimisation of this process is critical to ensure employers meet their legal duty of care to their workers, and ensure the likelihood of errors leading to significant financial and reputational loss is minimised.

One area of possible optimisation is considering how effectively such error checking tasks are allocated across a workforce of analysts. Automatic task allocation can be implemented by leveraging Artificial Intelligence (AI) to find solutions which allocate tasks to analysts with appropriate skills and experience (hence reducing completion time and risk of incompletion). For example, novel modelling techniques have been applied to personnel scheduling \citep{ERNST2004, van2013personnel}, task and team assignment for up-skilling workforces \citep{ahmadian2017job, starkey2016genetic}, selection policies to promote cross-learning to increase throughput \citep{nembhard2014selection}, and profiling worker skills to maximise efficiency in job assignment \citep{mourtzis2021intelligent, derkinderen2023optimizing}. While these approaches often address efficiency, which in-turn may reduce the average analyst's workhours, they do not sufficiently capture medium-term staff well-being objectives. Finding a solution should be multi-faceted; aiming to balance business orientated goals with those of the worker. 

A key gap in existing allocation and scheduling models, is therefore explicitly optimising worker well-being. Well-being objectives range from workload to task satisfaction, with the latter ensuring each analyst receives a fair share of rewarding tasks. Well-being objectives also include the autonomy of the worker and their individual preference for types of task they may like to complete. For financial service sector workforces, where high workload is often unavoidable, a critical aspect is to ensure that analysts continue to feel engaged and rewarded from their task set \cite{labor2022us}. Business orientated factors such as ensuring high priority tasks (i.e., those which are particularly time-dependent or most impactful to the NAV) are completed and maximising efficiency across the workforce, may often come at the cost of individual worker well-being factors such as task satisfaction. Financial service sector companies must strike a difficult balance between reducing risk from accurate NAV reporting, while maximising well-being objectives to keep voluntary staff turnover as low as possible. In addition to the disruption of training new workers, the cost of an analyst's job turning over is estimated at 50-125\% of the employee's annual salary highlighting the need for optimised well-being, even from a purely economic perspective \cite{forbes2018turnover}.

In this work, we formulate the task allocation and scheduling procedure as a multi-objective problem to quantify and balance both the needs of worker well-being and business orientated goals of efficiency and throughput. Using historical data, we compute expected task completion times by analyst to quantify worker suitability for an allocation based on skill. By considering this in relation to worker availability, we estimate the likelihood of each analyst completing their allocated tasks to assess business risk and the individual's workload. We characterise well-being based on both the likelihood of the allocated task being a true error (i.e., being a rewarding task), and the preference of the analyst (i.e., which tasks they find most satisfying). Pareto optimisation can then be found by our formal model which evaluates the fitness of a given allocation independently for each worker using these parameterisations of completion likelihood of allocated tasks, and well-being. 

In this context, finding sufficiently good solutions efficiently is critical. Implemented as a decision-support system for a human-in-the-loop, allocation solutions must be generated quickly for workforce managers who may then tweak allocation suggestions. Further, allocations may be re-evaluated as task lists and operational circumstances change through the day, again requiring efficiency. As a result, this problem can be framed as combinatorial optimisation where an exhaustive search (i.e., finding global optimal solution) is not plausible or necessarily required (given that managers may alter the recommended allocation). Therefore, specialised algorithms which can quickly rule out large parts of the search space, such as metaheuristics, are favoured.

One popular choice of metaheuristic, is the Genetic Algorithm (GA) which takes inspiration from evolutionary processes, such as mutation, crossover, and selection. GAs are a global optimisation technique, which utilise a population of candidate solutions across the search space. Evolutionary algorithms are popular choices for extensive multi-objective optimization problems as the set of solutions can explore the pareto front. For the purpose of our formal model, we utilise and validate a GA optimiser \citep{gad2021pygad} which aims to maximise the aggregated multi-objective fitness (or utility) across all analysts to ensure business orientated and worker well-being are balanced.

The key contributions of our work are to: 
\begin{enumerate}
    \item Develop a formal model with probabilistic objective functions (Section \ref{sec:objective_func}) for task completion probability and well-being (reward, preference). Our objectives optimise to ensure the completion of the highest importance tasks are prioritised in the allocation process. The model also contains heuristics which natively deal with pre-allocated and partially completed tasks, enabling the model to re-evaluate the solution as frequently as required, and emulate real-world scenarios. Our model natively handles different work schedules (e.g., part-time, full-time) for analysts. %Long tasks (i.e. those with completion time longer than worker availability) are automatically segmented. 
    
    \item Validate our formal model for the allocation and scheduling of a set tasks across a workforce. We validate the performance of the GA optimisation relative to baseline heuristics (Figure \ref{fig:fitness_hyperparameters}) and perform testing to provide suitable hyperparameters for realistic use cases (Section \ref{sec:hyperparameters}). 

    \item Consider the performance of our formal model in relation to current working practice (Figure \ref{fig:causality}). We simulate allocations from workforce managers and quantify the added value of of our formal model supported by a metaheurtic such as a Genetic Algorithm (Section \ref{sec:current_practice}).
    
    \item Evaluate the complexity scaling of single and multi-objectives problems for our formal model through asymptotic analysis. We further perform empirical analysis to understand the scaling of our formal model in typical real-world scenarios (Figure \ref{fig:scaling}).
\end{enumerate}

We validate our model using task and analyst data from a global top 10 asset manager taken over a two year period from 2020 to 2022. Due to data privacy, we present work here using simulated data drawn from typical working circumstances at the asset manager. In Section \ref{sec:lit_review} we provide further problem context about workforce planning, requirement for explicit optimisation of well-being, and metaheuristics. In Section \ref{sec:methods} we introduce the mathematical formulation of our allocation model and simulated data, before validating our models on a variety of test scenarios (Section \ref{sec:results}), comparing to baseline heuristics and current working practice. Finally we discuss the human-in-the-loop implementation of our model in Section \ref{sec:discussion} before concluding in Section \ref{sec:conclusion}.

\section{Literature Review} \label{sec:lit_review}
Increasing interest in automated approaches to workforce planning has provided a number of novel optimisation techniques aiming to improve allocation and scheduling of both work and workers. Reviews on personnel scheduling have defined taxonomies of areas within staff scheduling and rostering, often focused on increasing efficiency \citep{ERNST2004, van2013personnel}. Even approaches that aim to improve the skills of their workforce through task and team assignment \citep{ahmadian2017job, starkey2016genetic}, selection policies to promote cross-learning \citep{gomar2002assignment, nembhard2014selection} and profiling worker skills \citep{mourtzis2021intelligent, derkinderen2023optimizing} still optimise with long-term efficiency in mind. While up-skilling ensures workers develop sufficient skills to progress in their job and cross-learning ensures the overall workforce is better balanced, these approaches do not fully capture the objectives of worker satisfaction. Of particular relevance, we note \cite{shahbazi2019optimization} who develop a multi-objective formal model to allocate jobs in the construction industry optimising for both efficiency and career development. 

Business orientated goals of efficiency or throughput may come in direct competition with objectives to maximise worker well-being. Balancing multiple completing objectives has been previously studied in the context of scheduling \citep{cowling2006trade, quan2007searching}. Despite this, a current gap in knowledge are systematic task allocation methods which effectively balance the competing objectives of efficiency and well-being. Part of the problem could be due to difficulties in effectively modelling well-being. Contemporary theories of resource planning recognise that the alignment of needs and priorities of the business with those of those workers is critical for continued productivity \citep{demerouti2001job, bao2022job}. Despite this, even popular theories (e.g., job demand-resource model) only focus on relative workload and resourcing to define worker well-being \citep{shahbazi2019optimization}. 

Workforce planning is particularly challenging in the financial service sector due to the frequency of over-burden (over 40 hours as standard with junior analysts at Goldman Sachs reporting an average of 95 hours worked per week) \cite{goldman2022internal, labor2023financial}. Moreover, human resource management practices in finance are in stark contrast to contemporary theory. Analysts report significant impact to their physical and mental health due to working practices, with $>$75\% reporting work impacting their relationships with friends and family \cite{goldman2022internal, wso2021}. This practice does not recognise the two-way psychological relationship between employer and employee, championed by contemporary theory, outside of the legal and formal framework that an employee simply obtains work for fiscal reward \cite{bakker2017job}. Developing working practices (e.g., task allocation which optimises well-being) that better the `pyschological contract' between employee and employer enables workers to view their relationship as a two-way transaction, rather than one imposed against their interests \citep{demerouti2001job}.

To enable workforce planning which can optimise both for business orientated and well-being goals, it is critical to have an approach capable of finding solutions to extensive problems with multiple conflicting objectives or criteria. One option, Multiple-criteria decision making (MCDM), is a structured approach used to explicitly evaluate multiple conflicting criteria. MCDM typically takes a multi-step approach consisting of (i) problem formulation, (ii) objective definition, (iii) criteria selection, (iv) setting alternatives, (v) weighting the criteria and (vi) selecting an appropriate MCDM method. Approaches (collectively often referred to as multi-attribute decision making) include pair-wise comparison (e.g., Analytic hierachy process and analytic network process) \citep{saaty1988analytic, vincke1992multicriteria}, distance-based methods (e.g., TOPSIS \citep{tzeng2011multiple}), outranking methods (e.g., PROMETHEE \citep{brans1985note}) and value/utility function methods \citep{keeney1993decisions}. Despite success in resource allocation (e.g., land, water \citep{eastman1998multi, gebre2021multi}), problem formulation typically requires predetermined criteria weighting leading to a single solution. Another option is Reinforcement learning, where an agent learns policies to maximise the `reward function' or other user-provided reinforcement signals that accumulate from feedback in a dynamic environment. Here, the reward function would look to maximise scheduling efficiency while balancing user-provided input on task and work satisfaction. Despite popularity in the context of dynamic scheduling (see e.g., \citep{shyalika2020reinforcement}), continuous feedback on well-being is challenging.

Alternatively, multi-objective optimisation concerns optimising a set of objective functions simultaneously. Despite the positive characteristics of MCDM (e.g., tranparency of decision rules and more intuitive interpretation of outcomes) and reinforcement learning (e.g., adaptability and improvement over time), multi-objective optimisation does not require predetermined weights of each criterion leading to a single solution or continuous feedback on well-being. Instead, sets of solutions are generated in order to explore along a Pareto frontier enabling more detailed understanding between the trade-offs of conflicting objectives.

Given the operational bounds of our proposed model (i.e., requirement to be computationally efficient and only needing to find `good enough' solutions for a static problem), the most suitable subset of multi-objective optimisation techniques are metaheuristics. Metaheurstics are partial search algorithms designed to provide sufficiently good solutions with limited computational capacity and relative few assumptions made about the problem. Approaches include human mind inspired algorithms (e.g., rough set theory, fuzzy set theory, artificial neural networks \citep{pawlak1982rough, zimmermann2011fuzzy, russell2010artificial}, evolution inspired algorithms (e.g., genetic algorithms, evolutionary strategies \citep{holland1992adaptation, beyer2002evolution}) and swarm intelligence (e.g., ant colony, firefly, particle swarm optimisation \citep{dorigo1996ant, yang2010nature, kennedy1995particle}. 

In particular, Genetic Algorithms have been identified as a popular choice for scheduling and allocation related tasks with large, discrete, search spaces \citep{braun2001comparison}. GAs have been proven to be flexible, easy to implement and robust to non-linear and noisy objective functions while requiring minimal information about the search space (e.g., no gradient information required). For example, GAs have been utilised in scheduling for distributed computing systems \citep{vidyarthi2001maximizing, rekha2019efficient, zhou2020improved}, the deployment of unmanned aerial vehicles \citep{chen2003genetic, zhu2018multi, wu2021multi}, human-robot collaboration on tasks \citep{chen2012genetic, nunes2015multi, raatz2020task, liau2022genetic}, optimum resource planning for emergency departments \citep{yeh2007using, yousefi2018chaotic} and economics and business \citep{sieja2019use}. This background demonstrates the applicability of GAs to efficiently allocate a `batch' of tasks across the currently available workforce, finding a sufficiently good solution for workforce managers. We further discuss the suitability of other metaheuristics (e.g., ant colony, particle swarm optimisation) to find solutions for our multi-objective formal model in Section \ref{sec:discussion}.

\section{Methods} \label{sec:methods}
In this section, we outline the mathematical formulation of our task allocation model (Section \ref{sec:formal_model}), including definition of objective functions (Section \ref{sec:objective_func}), along the implementation of the genetic algorithm (Section \ref{sec:ga_methods}), and data used for model validation (Section \ref{sec:simulated_data}).

\subsection{Formal Allocation Model}\label{sec:formal_model}

In this section, we frame the problem of allocating tasks across a team of financial workforce of analysts as an optimisation problem. The aim of the optimisation algorithm is to allocate a set of tasks to maximize set objective(s) as defined below. Here, tasks relate to reviewing and resolving identified errors in trading data. These errors can result from human error, technology failures, market volatility and compliance issues. Furthermore, each task can be divided by type with each type having different expectations of completion time, required skills to review and resolve efficiently, and likelihood of being a true error. This is done to address the fact that analysts have a variety of skills and experience which affect their efficiency of completing different types of task. We furthermore assume that tasks are typically allocated in batches across a workforce for the given day. Although the optimisation is only done for a given batch, the model has been adapted to deal with pre-allocation and prior progress on tasks so can be re-run as task lists and operational circumstances change through the day.

\begin{table}[ht]
    \centering
    \begin{tabular}{c|c}
        Symbol & Meaning \\
        \hline
        $n$ & number of tasks\\
        $m$ & number of analysts\\
         $T=\{1,\ldots,n\}$ & set of tasks\\
         $A=\{1,\ldots,m\}$ & set of analysts\\
         $T_a \subseteq T$ & tasks allocated to analyst $a$\\
         $\theta_t \in \NP$ & task type\\
         $c_t \in \RP$ & task complexity\\
         %$d_t$ & task deadline or due date\\
         $\gamma_t \in [0,1]$ & probability of true positive\\
         $p$ & the number of priorities\\
          $\pi_t \in \{1,2,\ldots,p\}$ & priority of task (lower value is higher priority)\\
          $\eta_{a,\theta} \in \RP$ & analyst $a$'s efficiency of solving task of type $\theta$\\
          $\tau_a$  & analyst availability (in number of hours)\\
          $E(t,a)$ & expected execution time of task $t$ allocated to analyst $a$\\ 
          $\mu_{\theta}$ & average task duration of task of type $\theta$\\
          $\sigma^2_\theta$ & execution time variance of task of type $\theta$\\
          $U(T_1,\ldots,T_m)$ & Overall utility of an assignment\\ 
    \end{tabular}
    \caption{Notation summary}
    \label{tab:notation_summary}
\end{table}

In Table \ref{tab:notation_summary}, we present a summary of notation used in this article. In more detail, let $T=\{1,\ldots,n\}$ be the set of tasks to be allocated and $A=\{1,\ldots,m\}$ the set of (available) analysts in the workforce. Each task $t \in T$ is defined by a number of properties: $\theta_t \in \NP$ is the task type; $c_t \in \RP$ the task complexity; %$d_t$ the task deadline or due date; 
$\gamma_t \in [0,1]$ is the confidence of the task being a true positive (or, more generally, how interesting the task is); and $\pi_t \in \{1,2,\ldots,p\}$ is the priority (where a lower value means it is higher priority). If $c_t=1$ this means a task has average (or unknown) complexity, whereas $c_t>1$ and $c_t<1$ mean an above-average and below-average complexity respectively. 

Similarly, each analyst $a \in A$ has a number of properties: $\eta_{a,\theta} \in \RP$ is the analyst's efficiency of solving task of type $\theta$; and $\tau_a$ is the amount of time (number of seconds) that the analyst is available on the day in question. If $\eta_{a,\theta}=0$, this means the analyst is unable to execute the task of that type, $\eta_{a,\theta}=1$ means an average execution time, $\eta_{a,\theta}<1$ below average, and $\eta_{a,\theta}>1$ is above average.

A key part of the model is the objective function(s). Several objective functions will be defined below (Section \ref{sec:objective_func}). For now, we present the objective in general form. Let $T_a \subseteq T$ denote $a$'s assignment, i.e. a set of tasks allocated to analyst $a \in A$. Then, the tuple $\langle  T_1,\ldots,T_m \rangle$ denotes an assignment for each analyst. Furthermore, let $U(T_1,\ldots,T_m) \in \R$ denote the overall \emph{utility} or \emph{fitness} of a particular assignment. Given this, the aim is to maximise the assignment utility, which is given by:

\begin{equation}
\langle T^*_1,\ldots,T^*_m \rangle =\arg\max_{T_a \subseteq T, a \in A} U(T_1,\ldots,T_m)
\end{equation}
subject to each task being allocated to exactly one analyst, i.e., $\forall t \in T: t \in \cup_{a \in A} T_a$ and $\forall a1,a2 \in A, a1 \neq a2: T_{a1} \cap T_{a2}=\emptyset$

Another crucial aspect of the model is a task's execution time, which depends on both the properties of the task and the analyst. In particular, we assume the \emph{expected} execution time function, $E(t,a)$, is given by:

\begin{equation} \label{eq:exp_execution}
E(t,a)=\frac{\mu_{\theta_t} \cdot  c_t }{\eta_{a,\theta_t}}
\end{equation} 
where $\mu_{\theta_t}$ is the average execution time of a task with type $\theta_t$. Note that, in general, the execution time can also depend on the other tasks allocated. For example, if multiple tasks of the same type are allocated to the same analyst, these can often be completed more efficiently. However, this would make it challenging to estimate since it adds more unknown parameters. We leave this for future work. Finally, we assume that a task also has an execution time \emph{variance}, denoted by $\sigma^2_{\theta_t}$, indicating the variability in the execution time. For simplicity, we assume this is independent of the allocation.

\subsection{Objective Function}\label{sec:objective_func}

A key feature of the framework is that the objective function should be easily changed according to the specific application and client specifications, without having to fundamentally change the model or the algorithm to solve it. In other words, the system should be able to generalise to a wide range of objective functions. In addition, we are able to define \textit{individual} objective functions for every analyst. Depending on specific needs (i.e. different characteristics of individuals leading to increased job satisfaction) these can be tailored to the individual. Here we outline three objective functions appropriate for financial service sector domain, however, this can be easily modified to application-specific needs.

\subsubsection{Completion Probability}
From the business perspective, a reasonable objective is to maximise the throughput and explicitly optimise efficiency of task completion (e.g., see \citep{mourtzis2021intelligent, derkinderen2023optimizing}). An additional complication is that tasks have a sense of priority and it is critical that the highest priority tasks (i.e., those which are particularly time-dependent or most impactful to the NAV) are completed \citep{oncorps2024, funds-europe2024}. For those reasons, we aim to maximise the probability of completing tasks, given the analyst's availability, weighted by priority. Maximising completion probability naturally optimises for efficiency (i.e., finding analysts most efficient at a given task), while retaining fairness attributes. For example, framing as a probabilistic utility avoids scenarios where one particularly efficient worker is unfairly overloaded since the probability of them completing a task set tends to zero as more tasks are allocated (in comparison to minimising total completion time). This better aligns the needs and priorities of the business and workers when explicitly considering efficiency \citep{demerouti2001job, bao2022job}.

To calculate this, we assume tasks ($T_a^{\pi} \subseteq T_a$) are executed in order of priority ($\pi$) so that tasks in $T_a^{\pi}$ are executed before $T_a^{\pi'}$ if $\pi<\pi'$. To calculate the completion probability of a task, we assume execution times are normally and independently distributed with $\phi(x|\mu,\sigma^2)$ denoting the cumulative normal distribution. By defining $\mu^{\pi}_a=\sum_{t \in T_a^{\pi}} E(t,a)$ to be the expected total execution time of tasks, and $(\sigma_t^{\pi})^2=\sum_{t \in T_a^{\pi}} \sigma^2_{\theta_t}$ as the total variance, the probability of completing all tasks with priority $\pi$ is then given by:
\begin{equation}
Pr(\pi)=\phi\left(\tau_a \bigg| \sum_{\pi'=1}^{\pi} \mu^{\pi'}_a, \sum_{\pi'=1}^{\pi} (\sigma_t^{\pi'})^2\right)
\end{equation}
We can then write the conditional probability of completing tasks of priority $\pi$ given that tasks with $\pi-1$ (as well as all preceding tasks) are completed as:
\begin{equation}
Pr(\pi | \pi-1)=\frac{Pr(\pi)}{Pr(\pi-1)}
\end{equation}
noting that $Pr(\pi)=Pr(1) \prod_{\pi'=2}^\pi Pr(\pi' | \pi'-1)$. We also note that, given tasks are assumed to be completed in order, its means tasks of priority $\pi$ are completed only if tasks of priority $\pi - 1$ are already complete. This means that implicitly $Pr(\pi \cap \pi - 1) = Pr(\pi)$ (as denoted in the conditional probability. Now, in order to emphasize high priority tasks compared to lower priority tasks, we define the utility of an \emph{individual} analyst, denoted by $U_a^c(T_a)$ as follows:
\begin{equation} \label{eq:comp_prob}
U_a^c(T_a)=Pr(1) \prod_{\pi'=2}^\pi Pr\left(\pi' | \pi'-1\right)^{1/\pi'}
\end{equation}
Note that, by adding the power term $1/\pi'$ (conditional) probabilities of lower priority have reduced influence on the utility compared to those with higher priority. 
%Furthermore, to see why the product is more suitable than, for example, the sum, consider the following example. Consider a setting with $n$ identical analysts and $n \cdot k$ tasks. Assume all tasks have equal priority. Suppose that, if tasks are evenly allocated, the probability of any analyst completing the tasks is $0.80$. If all the tasks are allocated to a single analyst, the probability for that analyst is $\approx 0$ and is $1$ for remaining analysts. It is easy to see that, when $n$ is sufficiently large and when maximising the sum, the optimal solution is to allocate all tasks to a single analyst, which is an undesirable outcomes. Using the product this problem does not occur and tasks are allocated more fairly.

\subsubsection{Precision}
From the worker perspective, job satisfaction can be (partially) characterised in terms of workload (i.e., avoiding overburden and overtime) along with how engaging or rewarding a given set of tasks are (e.g., \citep{devaney2003job, sypniewska2014evaluation}). Completion likelihood aims to optimise to reduce potential for overtime but does not consider the fairness of the allocation in terms of rewarding tasks or what workers prefer to do. We now consider parameterisations for both task reward and worker preference. The former is an intrinsic property of the tasks with different types of task likely to be more rewarding, engaging, or fulfilling.

In the context of the financial service sector, the majority of tasks are error checking with different probabilities of identifying a true positive (rather than a false positive). While error checking tasks resulting in a false-alarm are likely to be perceived as repetitive and unrewarding, identifying true errors has tractable reward for the individual \citep{sypniewska2014evaluation}. Each task can be divided by type with each type having different expectations of completion time, required skills to review and resolve efficiently, and likelihood of being a true error, i.e. its \emph{precision} based on historical operations. Therefore, tasks can be scored based on their likelihood of reward (i.e., likelihood of being a true positive $\gamma_t$) and considered as an additional objective to optimise. Hence, the aim is to distribute the tasks in a fair manner so that no single analyst has too many false positives. 

Given that analysts can be allocated different number of tasks based on their availability, we propose to use the average $\gamma_t$ of the tasks allocated to analyst to determine their individual utility:
\begin{equation}
U^p_a(T_a)=\frac{1}{|T_a|}\sum_{t \in T_a} \gamma_t
\end{equation}
This simple objective function can be adjusted to optimise for various task associated scores aiming to increase well-being.  

\subsubsection{Task Preference}
Precision is a property associated with the task, and hence, optimisation can only ensure a fair balance of true-positives are distributed across the workforce. In addition, workers have individual preferences of the types of task they want to complete based on a mixture of personal skills, development goals, and enjoyment of the task itself. Therefore, a more personalised notion of task reward is to directly parameterise which tasks individual analysts \emph{prefer} to complete. In reality, a detailed characterisation of personalised reward (i.e., to explicitly quantify personal skills, development goals and enjoyment) would be a resource intensive procedure and outside the scope of the current work.

As a basic summarisation, we query analysts to respond to a Likert scale for each type of task. These scores are then normalised by worker (i.e., to correct for optimism or pessimism) and we look to explicitly maximise a time-weighted average of these preference scores. This is given by: 

\begin{equation}
U^t_a(T_a)=\frac{1}{|T_a|}\sum_{t \in T_a} \zeta_{t,a}
\end{equation}

where $\zeta_{t,a}$ is the preference score of a given task for the analyst.

\subsubsection{Overall Utility}
%$ Note that, for example, using the sum to compute the overall utility does not achieve the desired outcome. Consider a setting where all analysts are equivalent and all tasks are of the same duration, so all analysts get allocated an equal number of tasks. In this case, when taking the sum, the overall utility remains the same regardless of how the tasks are divided. On the other hand, the product again results in a more fair allocation. 
The above measures of individual utility can be combined by considering their product to balance business and worker orientated goals. Here, we define the business orientated utility to be given by equation \ref{eq:comp_prob} and the worker orientated utility to be the combination between precision and preference as follows:
\begin{equation}
U^w_a(T_a)=U^p_a(T_a) \cdot U^t_a(T_a) =  \frac{1}{|T_a|}\sum_{t \in T_a} \gamma_t \cdot \frac{1}{|T_a|}\sum_{t \in T_a} \zeta_{t,a}
\label{eq:worker_utility}
\end{equation}
Business and worker orientated utility can then be combined into:
\begin{equation}
U_a(T_a)=U^c_a(T_a) \cdot U^w_a(T_a) 
\label{eq:multi}
\end{equation}
We note that any combination of completion likelihood, individual preference and precision can be trivially defined. In words, this represents the expected completion time multiplied by the average satisfaction of the tasks (combining precision and preference). Then, to calculate the overall system objective, the individual utilities need to be combined into a single objective. Although the framework allows for different approaches, in this work we assume the overall utility is then computed by the \emph{product} of individual analyst utilities, i.e.:
\begin{equation}
U( T_1,\ldots,T_m )=\prod_{a \in A} U_a(T_a)
\label{eq:prod}
\end{equation}
Using the product is natural here since it indicates the probability of all analysts completing all tasks. In addition, in cooperative game theory literature, this type of joint utility is also known as the Nash product or Nash bargaining solution \cite{nash1950bargaining}, and has some attractive fairness properties. In particular, it is uniquely characterised by four properties or \emph{axioms}: Pareto efficiency, symmetry, independence of irrelevant alternatives and invariance to affine transformations of the utility function. The latter property means the solution does not rely on interpersonal utility comparisons, i.e. how each individual's utility is scaled. This is particularly important when there is no objective comparison such as money. For more detail, see e.g. \cite{binmore2014bargaining}. 

\subsection{Genetic Algorithm}\label{sec:ga_methods}
To efficiently find solutions to these objective functions we leverage a Genetic Algorithm (GA) implemented by PyGAD \citep{gad2021pygad}, an open-source Python library for GAs. Applied to the domain of task allocation, GAs obtain optimised solutions based on the fitness value (i.e. objective utility functions defined above) from a large pool of candidate solutions (population). A chromosome is a unique solution in the solution space with each chromosome containing $n$ (number of tasks) genes which can take $m$ (number of analysts) discrete values. The best solutions from each population are then passed to the next with variance induced by genetic operators such as mutation (gene alteration) and crossover (gene combination between parent solutions). PyGAD supports a variety of genetic operations and parent selection strategies, making it a flexible solution to allocation problems, and comprehensively identify appropriate hyperparameters (see Section \ref{sec:hyperparameters}). In Algorithm \ref{algorithm}, we present a pseudocode of our proposed formal model including the GA procedure.

\begin{algorithm}
\begin{algorithmic}
\State START: PREPROCESS($\tau, \mu$)
\State// \textbf{Evaluate Problem}
\State Compute total sum of workforce availability $\tau_{total}$;
\State Compute total sum of expected task duration $\mu_{total}$;
\State Identify tasks with prior allocation and completion and adjust $\tau_{total}$ and $\mu_{total}$ appropriately;
\State
\State// \textbf{Select Task List}
\State If $\tau_{total} >> \mu_{total}$  drop lowest priority tasks;
\State 
\State START: Find solution with GA($j, \chi, \eta$)
\State// \textbf{Initialise Generation}
\State $k := 0$; 
\State $P_k$ := population of $j$ randomly generated solutions
\State// \textbf{Evaluate $P_k$}
\State Compute fitness(i) for each $i \in P_k$;
\State// \textbf{Iterate to find best possible solution in 50 generations}
\While{$k < 50$}
\State// \textbf{Create generation k + 1}
\State// \textbf{1. Elitism}
\State Select $(1 – \chi) \times j$ members of $P_k$ and insert into $P_{k+1}$;
\State 
\State // \textbf{2. Crossover}
\State Select $\chi \times j$ members of $P_k$ (steady state); 
\State pair them up and produce offspring; 
\State insert the offspring into $P_{k+1}$;
\State
\State // \textbf{3. Mutate}
\State Random mutation applied at gene level;
\State High adaptive mutation rate for $\eta \times j$ members with lowest fitness
\State Low adaptive mutation rate for $(1 - \eta) \times j$ members with highest fitness
\State
\State // \textbf{Evaluate $P_{k+1}$}:
\State Compute fitness(i) for each $i \in P_k$;
\State 
\State // \textbf{Increment:}
\State k := k + 1
\EndWhile
\end{algorithmic}
\caption{Pseudocode of the Formal Model with GA generated solutions.}\label{algorithm}
\end{algorithm}

\subsection{Simulation} \label{sec:simulated_data}
\begin{figure}[ht]
    \centering
    \includegraphics[width=\textwidth]{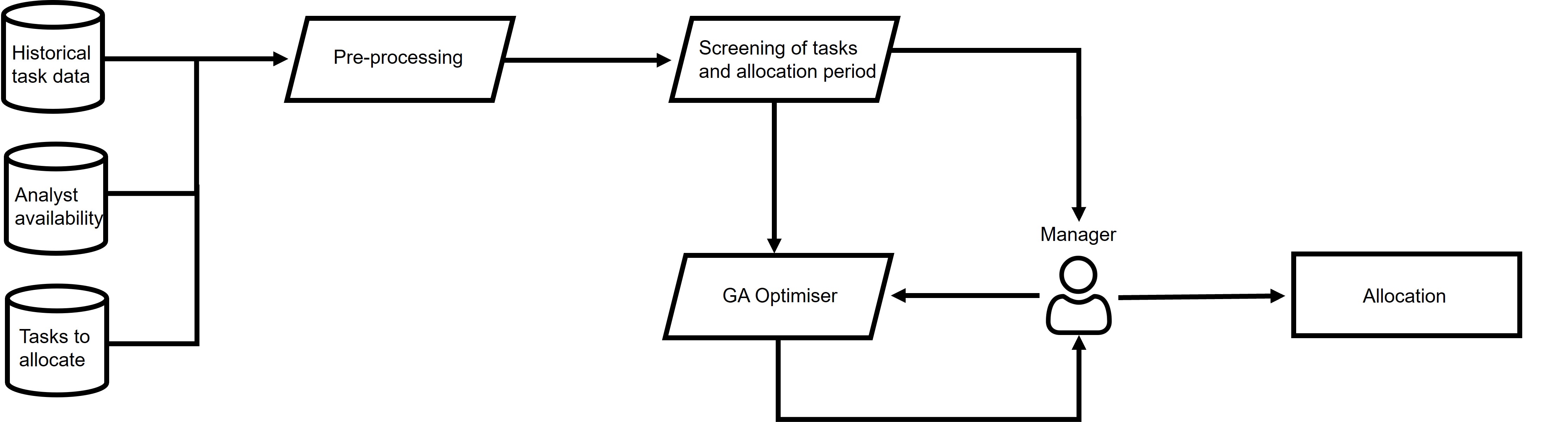}
    Source: Authors. 
    \caption{Flow diagram showing the working context of our proposed model. Historical data on previous tasks and analysts is first combined with data on the current allocation period. This is pre-processed to estimate parameters for the current allocation period (e.g., expected completion type of tasks to be completed). The task list and analyst availability is screened to identify significant overall burden (i.e., expected total completion time greatly exceeds availability) or tasks which are individually too long to complete in the period. Any warnings are sent to the manager who can decide to remove or segment tasks if necessary. This updated task list is then initially allocated by the GA optimiser, which the workforce manager reviews and amends before passing to workforce. }
    \label{fig:flow_diagram}
\end{figure}

Throughout model development real-world task and workforce data from a global top-10 asset manager was utilised, which this algorithm will service. In Figure \ref{fig:flow_diagram} we show how our algorithm fits into the workflow of task allocation overseen by a manager. Historical task data is combined with data of the current task list and available analysts, before pre-processing and screening (i.e., identifying if the expected total completion time greatly exceeds availability). Any warnings are sent to the manager who can decide to remove or segment tasks if necessary. This updated task list is then initially allocated by the GA optimiser, which the workforce manager reviews and amends before passing to workforce.

Due to privacy concerns, we publicly evaluate our model using simulated data drawn from typical operating circumstances at the asset manager. To generate estimates of task complexity in Table \ref{tab:task_by_type} we define 5 typical types of tasks with varying expected completion times and frequency of occurrence. To simulate test scenarios, we randomly generate a list of tasks with occurrence frequency and expected completion times drawn from this table. We assume task duration and variance by type to be normally distributed.

\begin{table}[ht]
    \centering
    \begin{tabular}{c|c|c|c}
        Type, $\theta_t$ & $\mu_{\theta}$ (seconds) & $\sigma^2_\theta$ (seconds) & Relative frequency \\
        \hline
        $A_t$ & 1800 & 90000 & 1\\
        $B_t$ & 3600 & 810000 & 0.75\\
        $C_t$ & 7200 & 1440000 & 0.5\\
        $D_t$ & 14400 & 7290000	& 0.25\\
        $E_t$ & 21600 & 12960000 & 0.05\\
    \end{tabular}
    \caption{Properties of five tasks types which are used as a basis to simulate real-world scenarios. Mean duration ($\mu_{\theta}$) and variance ($\sigma^2_\theta$) are the expected global completion (and variance) time for a task of that type across all analysts. Relative frequency defines the proportion of each task type represented in a simulated scenario.}
    \label{tab:task_by_type}
\end{table}

To calculate the expected execution time function given in equation \ref{eq:exp_execution} and emulate the differences in skills of analysts we modify $\eta_{a,\theta}$ (efficiency by task type) randomly by a factor between 0.9 - 1.1 when generating a simulated analyst. We note that setting uniform analyst efficiencies does not impact the ability of the model to converge. In practice analysts will also balance newly allocated tasks with left-over work from the prior workday. To emulate this, we pre-allocate 5\% of our simulated tasks with a randomly generated prior progress time.

To reflect the heavy workload found in asset management, we define test scenarios that are `difficult but achievable'. Practically this means defining a task list which is expected to between 1.01 and 1.1 times the total analyst availability. Due to the nature of a probabilistic completion time fitness function, complete over-burden can result in zero probability for an analyst and the algorithm failing to converge. For this reason, problems (or task lists) should be pre-screened to ensure they meet this criterion. For problems with low workload, we heavily penalise allocations with analysts that have zero allocated tasks to ensure a fair distribution is still found.

\section{Results} \label{sec:results}

This section demonstrates the performance of our allocation model with several GA hyperparameter set-ups relative to baseline allocation algorithms. We also consider the run-time complexity of our approach and further evaluate our formal model on a set of realistic use cases of different size and optimising goals (i.e. fitness function). For comparison between hyperparameter choices and to baseline algorithms, we simulate a basic test scenario of 65 tasks to be allocated between 10 analysts to optimise for priority weighted completion time. To understand how our model scales, we adopt Big $\mathcal{O}$ notation to describe the limiting (worst case) behaviour. We further simulate test scenarios up-to a size of 325 tasks and 50 analysts for both single (completion probability only) and multi (both completion probability and true positive probability) objective functions to evaluate empirically.

\subsection{Hyperparameter Selection} \label{sec:hyperparameters}

To select an effective combination of GA hyperparameters we validate performance for a variety of population sizes along with different parent selection, mutation, and cross-over techniques. In Figure 1 (left panel) we show the average performance (as defined by total utility) of three distinct hyperparameter approaches; green: steady state parent selection with adaptive mutation, blue: tournament parent selection with adaptive mutation; red: steady state parent selection with scramble mutation. All other hyperparameters are consistent as defined in Table \ref{tab:best_hypers}. Our key findings are as follows:
\begin{itemize}
    \item \textbf{Population size and number of generations}: 50 generations with 500 solutions per population reliably result in a converged solution (as shown by Figure 1 right panel). Increasing the number of generations (at the cost of population size) generally results in decreased performance, however effect is minimal. 
    
    \item \textbf{Mutation}: Adaptive mutation results in the best performance. As outlined in \cite{marsili2000adaptive}, the weak point of `classical' GAs is that mutation is randomly applied to all chromosomes, irrespective of their fitness. Adaptive mutation solves this by applying a high (low) mutation rate to low (high) fitness solutions overcoming the usual trade-off between making incremental improvements and finding better maxima. Scramble mutation (blue) performs particularly poorly since it keeps the number of tasks assigned per analyst constant (similar for inversion and swap mutation).
    
    \item \textbf{Parent Selection}: Steady State (green) outperforms Tournament parent selection (blue). Along with defined elitism, there may be minimal selection of low fitness solutions to be passed to the next generation, highlighting the importance of being able to define a high mutation rate and escape local maxima.
    
    \item \textbf{Crossover}: Crossover choice had minimal effect on performance for this use case, leading to single point crossover being selected.
\end{itemize}

\begin{figure}[ht]
    \centering
    \includegraphics[width=\textwidth]{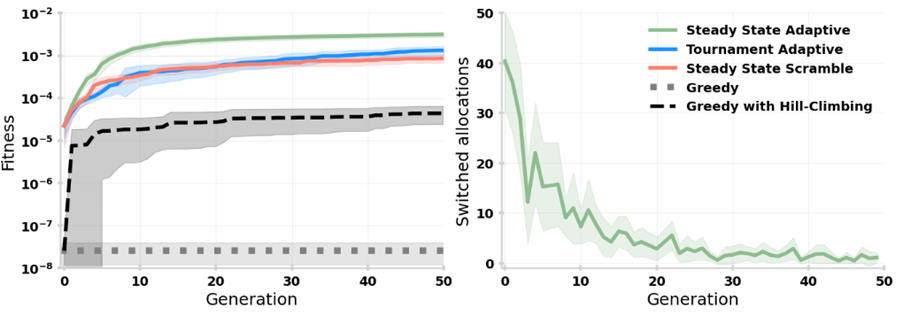}
    \caption{(Left) Average performance (as defined by fitness of best solution) of genetic algorithms (solid lines; green, blue, red) and baseline algorithms (greedy – dotted grey, greedy and hill-climbing – dashed black). The average is found by performing 20 runs of each algorithm set-up and the shaded regions represent the 95\% confidence interval. (Right) Average number of tasks switching allocation per generation for best solution. Only shown for best GA hyperparameter set-up (green - Table \ref{tab:best_hypers}).}
    \label{fig:fitness_hyperparameters}
\end{figure}

Our final hyperparameters are described in Table \ref{tab:best_hypers} (represented by the green line in Figure \ref{fig:fitness_hyperparameters}). For larger scale problems, we utilise the same hyperparameters however increasing the population size, number of parents mating and elitism proportionally to the number of tasks. 

\begin{table}[ht]
    \centering
    \begin{tabular}{c|c}
        Population Size & 500\\
        Generations	& 50\\
        Crossover Type & Single Point\\
        Mutation Type & Adaptive (Random with variable rates)\\
        Mutation Probability & [0.9, 0.05]\\
        Parents Mating & 50\\
        Elitism	& 10\\  
    \end{tabular}
    \caption{Hyperparameters for fiducial GA run validated on basic test scenario of 65 tasks allocated across 10 analysts.}
    \label{tab:best_hypers}
\end{table}

\subsection{Comparison to baseline heuristics}

To validate our choice of a metaheuristic such as a Genetic Algorithm for efficiently finding allocation solutions to our formal model, in Figure \ref{fig:fitness_hyperparameters} (left panel) we directly compare to baseline heuristics. 

We adopt a greedy allocation policy \citep{dijkstra2022note} (grey dotted line), which allocates sequentially by task to the analyst which maximises global utility at each step. Generally, a greedy policy in optimisation problems does not produce an optimal solution, however can yield local maxima with reasonable run-times. Despite being computationally efficient, this typically provides a poor scoring solution and often fails when applied to more complex fitness functions. We improve upon the initial greedy allocations through application of a hill climbing policy (black dashed line). Hill climbing is an iterative search optimisation technique which starts with an arbitrary solution, then makes incremental changes to find better local solutions. Here, the hill climbing policy randomly swaps two task allocations to find an improved solution (from the initial greedy solution). The GA optimisation significantly outperforms both the baseline greedy algorithm and the hill climbing improvement (scaled to be same number of utility function evaluations as the GA).

\subsection{Stopping condition}
To validate an appropriate stopping condition (or number of generations) for our GA, we consider both the convergence of fitness score (Figure \ref{fig:fitness_hyperparameters} left panel) and the number of task allocations changed between best solutions of each generation (right panel). We note that fitness (and changes to best allocation) changes rapidly in the first 10 generations, settling towards generation 20 and showing steady but minimal improvement (and changes to best allocation) towards generation 50. We find 50 generations is an appropriate stopping point to ensure a high fitness solution is consistently found, however, if a quicker compute time was required, solutions from generation 20 onwards could be used. 

\subsection{Scalability}

We now consider the run-time complexity of our formal model supported by a GA optimiser. We adopt Big $\mathcal{O}$ notation to describe the limiting (worst case) behaviour of our approach as $n \to \infty$ (where $n$ is the size of the population in the GA). The key operations and time-complexity are as follows: i) initialisation of population: $\mathcal{O}(n)$, ii) evaluation of utility: $\mathcal{O}(n^{2})$ (since utility function evaluation scales with $\mathcal{O}(n)$ and must be evaluated for each worker whose number we assume scales with the number of tasks and hence population size $n$), iii) selection $\mathcal{O}(nlogn)$ (requires sorting), iv) crossover $\mathcal{O}(n)$, v) mutation $\mathcal{O}(n)$ and vi) termination conditions $\mathcal{O}(1)$ (since is manually defined to be 50 generations). The dominating term is therefore $\mathcal{O}(n^{2})$.

\begin{figure}[ht]
    \centering
    \includegraphics[width=\textwidth]{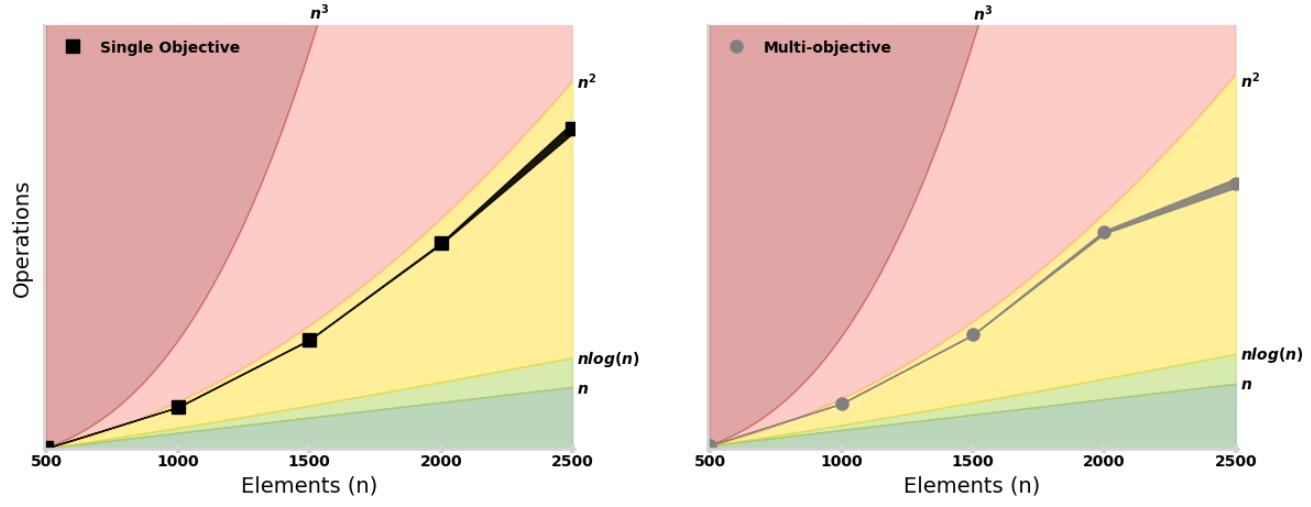}
    \caption{Run-time complexity for single objective (left) and multi-objective (right) allocations for a range of problem sizes, as denoted by the GA population size ($n$). Connecting lines show the width of the 95\% CI. The background is shaded by scaling complexity (relative to smallest scale problem in each panel; i.e. 65 tasks and 10 analysts) divided by $n$, $nlog(n)$, $n^2$, $n^3$ (in order of increasingly poor scaling).}
    \label{fig:scaling}
\end{figure}

In Figure \ref{fig:scaling}, we also empirically estimate the run-time scaling for our allocation model for a set of realistic scenarios with increasing numbers of tasks and analysts for single (black) and multi (i.e, completion likelihood and true positive likelihood; grey) objective problems. The absolute run-time units are seconds (based on serial evaluation on a single node cluster with 32GB of memory). In response to the increasing problem scale, we increase the population size of our GA search space proportionally to the number of tasks to allocate (i.e. double number of tasks equates to double the population size). Despite this linear scaling, the number of fitness function evaluations also increase due to a greater pool of analysts. We fix the number of generations to be 50 for each problem size, finding solutions to be well converged. For a given problem size, increasing the number of generations would lead to a linear increase in computation time. This is in-keeping with our theoretical complexity scaling limit of $n^{2}$.

% Scability is considered by normalising with respect to the run-time of the smallest problem scale. This intermediate run-time scalability reflects the scaling of the search space and highlights the challenges of applying our pipeline to significantly larger problems. It should be noted that baseline algorithms (greedy and hill-climbing) scale significantly worse and frequently fail completely for multi-objective optimisations. While the implementation of PyGAD enables multi-processing and threading, we find no significant differences in run-times relative to the serial runs presented here.

We note that this holds one key assumption that solutions converge consistently when increasing the scale of the GA search space linearly with the number of tasks to be allocated. In reality for larger use cases, it would be more suitable to scale the population size proportional to both the increase in workers and tasks, therefore our formal model would instead scale $\mathcal{O}(n_{task}^{3})$ (where here $n_{task}$ is the number of tasks to allocate).

\subsection{Comparison to current practice} \label{sec:current_practice}

\begin{figure}[ht]
    \centering
    \includegraphics[width=\textwidth]{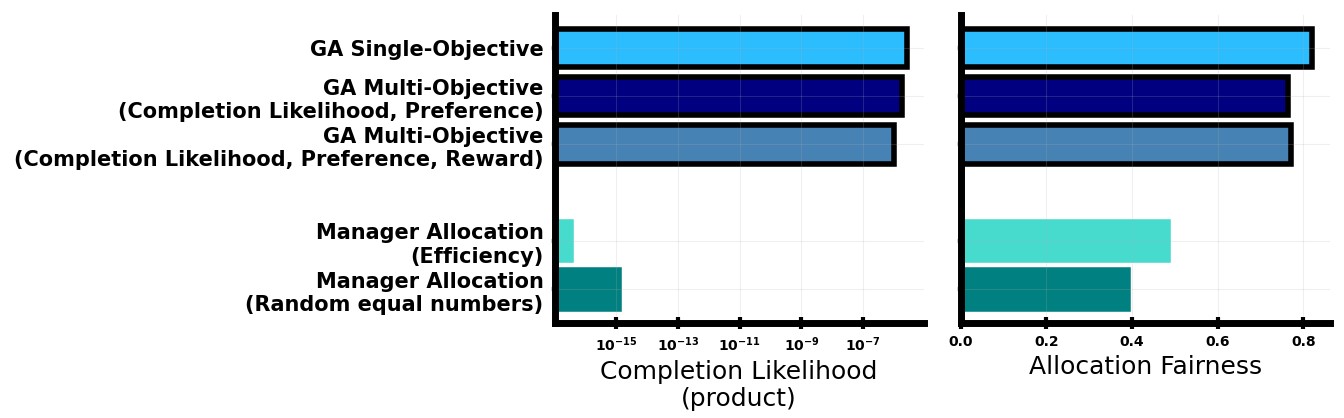}
    \caption{Comparison between simulated manager-led allocations and the proposed formal model. In each panel a simulated manager allocating to maximise efficiency (light green), and, to balance task numbers (dark green) is compared to a single-objective GA to optimise for completion likelihood (light blue), a multi-objective GA optimising for completion likelihood and task preference (dark blue) and a multi-objective GA optimising for completion likelihood, task preference and reward (grey blue). The scores for GAs solutions are found from the average of 5 runs. (Left) Completion likelihood across complete workforce (i.e., product of individual completion likelihoods for their allocation). (Right) Allocation fairness. This is maximum difference between the best and the worst allocation by analyst. A given analyst's allocation is evaluated by the product of completion likelihood, the time-weighted preference score and the time-weighted reward (anomaly score).} 
    \label{fig:causality}
\end{figure}

We now evaluate our formal model relative to typical operating circumstances (i.e., task allocation performed by workforce manager) to assess its real-world benefit. To emulate how a manager may allocate tasks across a typical workforce we consider two manager strategies:
\begin{enumerate}
    \item \textbf{Efficiency}. Here the manager allocates tasks one-by-one (in order of priority) to the analyst who has historically performed that type of task most efficiently. Once a given analyst is over-burdened, following tasks are then allocated to the second most efficient analyst, and so on. This is effectively the greedy algorithm approach described above. This allocation strategy also is designed to minimise potential impact to the NAV by allocating the highest priority (and hence highest materiality) tasks one-by-one.
    \item \textbf{Balancing task numbers}. Here the manager allocates (pseudo-randomly) to ensure each analyst has roughly the same number of tasks, a basic notion of fairness. This approach ignores other measures of reward, the overall length of tasks, and the likelihood of each analyst completing their allocation.
\end{enumerate}
In Figure \ref{fig:causality}, we compare the simulated manager allocations against our single and multi-objective GA approaches. Scores for the GA approaches are found from the average of 5 runs. Here we use a test scenario with 65 tasks allocated across a workforce of 10 analysts (as described in Section \ref{sec:hyperparameters}). For each task we generate a `reward' score based on the anomaly score of the task. For each analyst we also generate `preference' scores by task-type. Each GA optimises completion likelihood along with preference (navy) and both preference and reward (grey blue).

In the left panel we quantify the success of simulated managers and GAs at maximising the likelihood of completing the complete set of tasks. This is effectively the business orientated goal. This is found by the product of completion likelihood across all 10 analysts (i.e., equation \ref{eq:prod}). As discussed in Section \ref{sec:objective_func}, we use the product rather than the average or sum to ensure fairness in evaluation. We find all GA approaches offer significant improvement over the simulated manager allocations, increasing global likelihood of tasks being completed by at-least 7 orders of magnitude. In particular we note the poor performance of the simulated manager allocating tasks to the most efficient analyst; highlighting the pitfalls of greedy policies (even relative to a pseudo-random allocation).

% \red{In the middle panel we quantify the success of simulated managers and GAs at maximising well-being across the workforce. For each analyst the average (weighted by time) preference score for their allocated task list is calculated. The product across the 10 analysts is then found. The multi-objective GA performs best at allocating to maximise well-being, as expected due to it explicitly optimising preference scores. We note that approaches explicitly allocating to optimise completion likelihood (i.e., manager efficiency and single-objective GA) come at the cost of significantly lower well-being scores relative to a pseudo-random allocation (dark green). This highlights the need for multi-objective approaches to fully balance business orientated and worker well-being goals, which can be difficult for managers to consider both (and algorithms which do not consider well-being).}

In the right panel we quantify the overall fairness of the allocation across the workforce. For each analyst the `quality' of their allocation is found by the product of the completion likelihood, the (time-weighted) average preference score, and the (time-weighted) average anomaly score (reward). This score balances both worker over-load and reward of tasks. The fairness of each allocation is then found by finding the difference between the best and the worst scoring allocation by-analyst. We find that both single and multi-objective GA approaches result in significantly fairer allocations relative to the simulated manager allocations.

\section{Discussion} \label{sec:discussion}
In practice, AI powered allocation and scheduling recommendations will be used with resource managers in-the-loop (i.e., human-in-the-loop; HITL). HITL integrates human judgement and expertise into the decision making process, augmenting the capabilities of decision-support tools, such as our proposed formal model. Human involvement in the decision-making process fosters trust and transparency, as ultimately the resource manager has final say. Another critical factor, is the contextual domain knowledge of resource managers, often difficult to capture in modelling. Practically this AI recommendation would therefore be used as a starting point for the manager to `tweak' depending on current operational circumstances and personal domain knowledge (e.g., escalating certain tasks to be more critical than identified in the system).

\begin{figure}[ht]
    \centering
    \includegraphics[width=\textwidth]{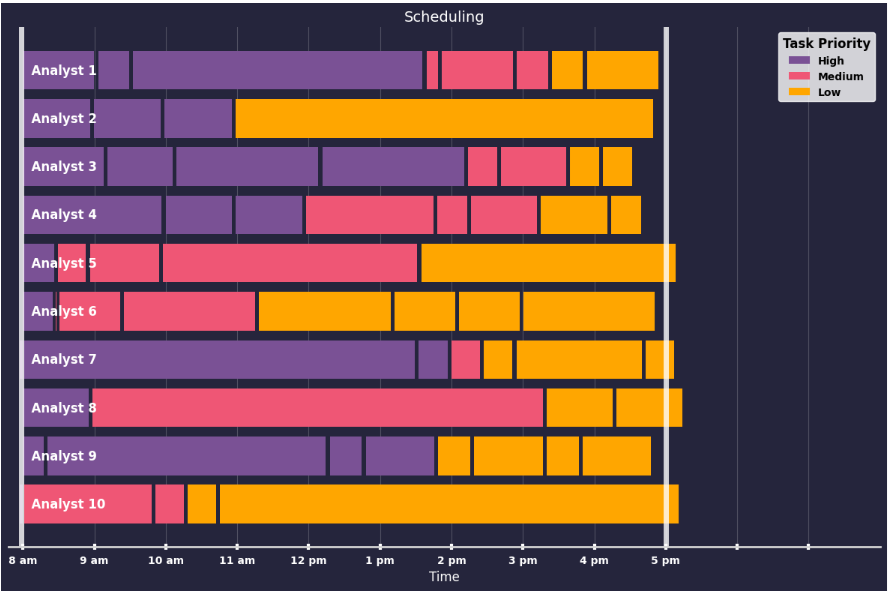}
    \caption{Example task allocation and scheduling found from the multi-objective test case with 65 tasks and 10 analysts. Each row shows the allocation for a given analyst throughout their given workday. Tasks are coloured by priority (purple: high, pink: medium, orange: low). The available workhours are given by the white solid lines. Note: work breaks are not explicitly drawn onto the schedule with expected completion times compensating for breaks through the day.}
    \label{fig:allocation}
\end{figure}

In Figure \ref{fig:allocation}, we show a visualisation containing an example allocation found from our GA model applied to the multi-objective test scenario with 65 tasks and 10 analysts. In this instance, a manager may identify high priority tasks which are scheduled to be completed later in the day (e.g., Analyst 3's 4th task) and choose to re-allocate to ensure they are completed in a timely manner (e.g., swap with Analyst 10's first task). While HITL likely reduces the utility score of an allocation, it still offers the best approach to maximise business and well-being (significantly outscoring current working practice), while ensuring the hybrid approach is robust by avoiding pitfalls not captured by modelling. Additionally, human input may enable soft and hard constraints on allocation to be identified (e.g.,  that a given analyst cannot a specific task due to external factors not previously captured in data). Constrained optimisation (and hence narrowing of the search space) potential enables solutions to be found faster.

Another important purpose of the decision-support system is to identify significant pressures on the workforce (i.e., when even optimised solutions may be expected to take significantly longer than the total available workhours for a team). Providing insights into the available allocations, along with specifics about the severity of the potential overload (or underload) is therefore critical to inform business processes (e.g. whether the situation can be solved by over-time or severe enough to consider bringing in analysts from external teams). 

An important adaptation to our formal model, is the ability to consider tasks which are pre-allocated or have prior completion. As a result, we are able to periodically (or on-demand) re-evaluate the allocation and provide updated recommendations. Throughout the day progress will naturally fluctuate by analyst relative to initial expectations and operational circumstances may change. Therefore, the ability to efficiently re-optimise task allocations to improve business and well-being needs is critical. Additionally if significant pressures are identified and external members are brought in to help, again, quick re-allocation is required. 

In this manuscript, we have demonstrated one example of a metaheuristic (i.e., GA) can quickly converge on solutions with significant improvements on current working practice. Combinatorial optimisation with multiple objectives makes GA a popular choice, however, we note the validity of other approaches such as swarm intelligence (e.g., firefly, particle swarm optimisation: PSO) which have shown increasing popularity in recent years. In the instance of PSO, we note one potential advantage is the ability for faster convergence and more diversity in search trajectories (due to momentum effects) \citep{goel2020extensive}. A promising variant of GAs for faster convergence have been those with adaptive mutation and crossover rates (e.g., \citep{han2022improved}). Here, we demonstrate taking an adaptive mutation strategy (dependent on the fitness of the solution) results in faster convergence on higher quality solutions relative to other strategies (e.g., scramble). We look to future work to compare the performance of other metaheuristics utilising our formal model.

Finally, we demonstrate that a formal model explicitly optimising for well-being can still find solutions which still significantly improve efficiency relative to current working practice (Figure \ref{fig:causality}). Even with potentially conflicting goals, this approach recognises the two-way psychological relationship between employer and employee, outside of the legal and formal framework that an employee simply obtains work for fiscal reward \cite{bakker2017job}. Developing working practices that better the `pyschological contract' between employee and employer (e.g., allowing workers to define preferred tasks) enables workers to view their relationship as a two-way transaction, rather than one imposed against their interests \citep{demerouti2001job}. In turn, finding solutions which improve efficiency benefit both the business and workers by reducing the potential overburden of overtime. In future work we look to develop more detailed mathematical representations of well-being which quantify factors such as short-term satisfaction and personal skill development.

\section{Conclusions} \label{sec:conclusion}
Financial service sector companies routinely check and resolve errors across huge data sets, putting significant pressure on workforces with increasingly manual tasks. Automation is critical to ensure tasks are allocated efficiently and fairly. In this article we introduced a formal model leveraging a GA to allocate and schedule tasks across a workforce. Through definition of objective functions for completion probability and worker reward, we present a methodology to allocate tasks to both maximise business orientated goals and worker well-being. The GA significantly outperforms baseline heuristics and scales intermediately to a range of typical working circumstances, demonstrating the potential for metaheuristic approaches. This formal model is intended to service workforces at a global top 10 asset manager. Finally, we note the applicability of our formal model to other domains, in particular medical triage within an emergency department which we look to in future work.


 \bibliographystyle{elsarticle-num} 
 \bibliography{cas-refs}





% \end{thebibliography}
\end{document}